\documentclass[showpacs,prb,twocolumn,floats]{revtex4}
\usepackage{graphicx}
\usepackage{amsmath}
\usepackage{amssymb}
\usepackage{epstopdf}
\usepackage{color}

\newcommand{\e}{\mathrm{e}}

\renewcommand{\Re}{\operatorname{Re}}
\renewcommand{\Im}{\operatorname{Im}}

\begin{document}
\title{Electronic states in a graphene flake strained by a Gaussian bump}

\author{D.~Moldovan}\email{dean.moldovan@ua.ac.be}
\author{M.~Ramezani Masir}\email{mrmphys@gmail.com}
\author{F.~M.~Peeters}\email{francois.peeters@ua.ac.be}
\affiliation{Departement Fysica, Universiteit Antwerpen \\
Groenenborgerlaan 171, B-2020 Antwerpen, Belgium}

\begin{abstract}
  The effect of strain in graphene is usually modeled by a pseudo-magnetic vector potential which is, however, derived in the limit of small strain. In realistic cases deviations are expected in view of graphene's very high strain tolerance, which can be up to $25\%$. Here we investigate the pseudo-magnetic field generated by a Gaussian bump and we show that it exhibits significant differences with numerical tight-binding results. Furthermore, we calculate the electronic states in the strained region for a hexagon shaped flake with armchair edges. We find that the six-fold symmetry of the wave functions inside the Gaussian bump is directly related to the different effect of strain along the fundamental directions of graphene: zigzag and armchair. Low energy electrons are strongly confined in the armchair directions and are localized on the carbon atoms of a single sublattice.
\end{abstract}


\maketitle

\twocolumngrid

\section{Introduction}

A single layer of carbon atoms called graphene has become a very active field of research in nanophysics\cite{no,zh}. Graphene has excellent electrical and thermal properties, e.g. massless and chiral Dirac fermions which move with a Fermi velocity of about 1/300 the speed of light, a linear spectrum close to the $K$ and $K'$ points\cite{no,zheng} of the Brillouin zone, anomalous integer quantum Hall effect in the presence of a magnetic field, the Klein paradox, i.e. unusual high transmission when electrons pass classically forbidden regions,  Aharonov-Bohm effect in graphene rings, extraordinary stiffness, unexpected mechanical properties, and thermo-mechanical and electronic properties that are highly affected by external particles and dopants. These properties of graphene have attracted considerable attention and make it a promising material for future electronic and opto-electric devices.

\begin{figure}[t]
  \centering
  \includegraphics[width=8.6cm]{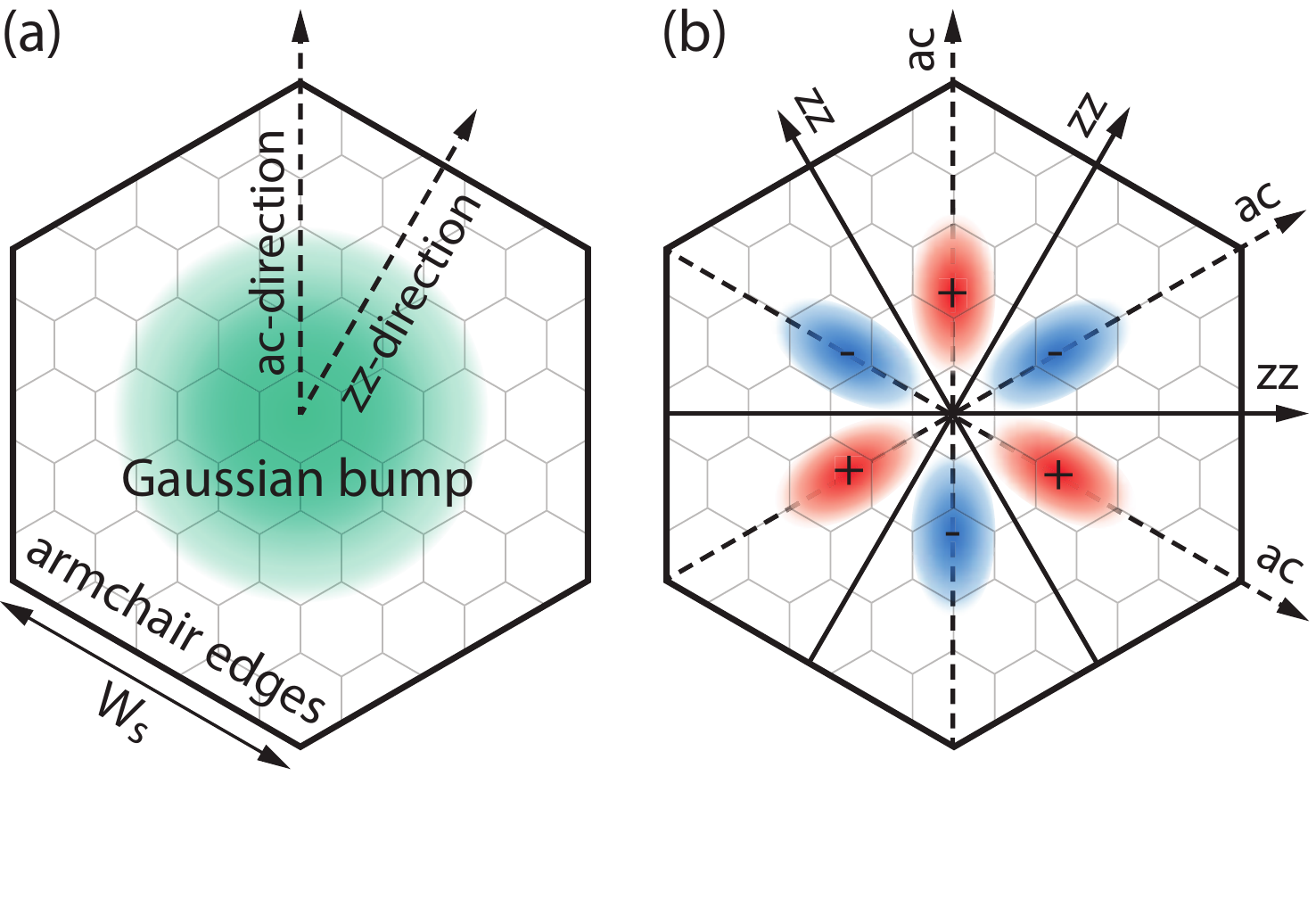}
  \caption{(Color online) (a) Hexagon shaped graphene flake with armchair edges strained by a Gaussian bump in the center. The dashed arrows show the armchair and zigzag directions in the radial direction of the bump. (b) The bump-generated pseudo-magnetic field, as calculated from the traditional form of the pseudo-vector potential. Red (blue) color corresponds to positive (negative) magnetic field. }
  \label{fig:Intro_1}
\end{figure}

An interesting recent prediction is that a geometrical deformation of the graphene lattice results in local strain that acts as a pseudo-magnetic field on the electronic degrees of freedom and which leads to a pseudo-quantum Hall effect \cite{M4}. Deformation due to elastic strain changes the hopping amplitude of the carbon atoms and induces an effective vector potential that shifts the Dirac points\cite{M5}. With a proper geometrical deformation it is possible to create large pseudo-magnetic fields which can reach up to several hundreds of Tesla\cite{M1, M6}. Over the last few years much effort has been devoted to find ways of controlling graphene's electronic properties by strain. Applying in-plane strain with triangular symmetry has been shown\cite{M4} theoretically to result in an uniform pseudo-magnetic field of the order of $10$ T. It was reported experimentally\cite{M12} that nanobubbles grown on a Pt(111) surface induce pseudo-magnetic fields of more than $300$ T. Landau quantization of the electronic spectrum was observed by scanning tunneling microscopy. Thus, with such large strain-induced pseudo-magnetic fields, one can control the electronic properties of graphene through strain engineering\cite{M11}. Recently, it was shown experimentally that an external nonuniform electric field is able to induce local deformations in graphene with different curved shapes \cite{M15} and thus one should be able to induce a pseudo-magnetic field through an electric field.

In this paper we investigate the different effects that are induced by inhomogeneous strain in graphene. We consider a hexagon shaped graphene flake that is strained out-of-plane by a Gaussian bump placed in its center. The effects of strain in graphene can be modeled using a pseudo-magnetic vector potential. In the case of a Gaussian bump, the traditional form of this vector potential\cite{Suzuura2002} results in a three-fold symmetric pseudo-magnetic field, as illustrated in Fig. \ref{fig:Intro_1}(b). Recently, it has been shown in Ref. \onlinecite{Kitt2012} that additional lattice corrections are required in order to accurately calculate the pseudo-magnetic vector potential. However, these strain-induced lattice vector corrections do not contribute to the pseudo-magnetic field and may be neglected\cite{Kitt2012e,Juan2013}. Only the strain induced hopping parameter changes will affect the intensity of the pseudo-magnetic field, but this is generally derived only up to first order in strain. Given graphene's excellent mechanical properties, it can sustain strain up to $25\%$\cite{Lee2008}. At that point strain can no longer be considered to be small. For this reason, we investigate additional corrections to the vector potential to higher order in the strain and we compare this pseudo-magnetic field model to results obtained with the full tight-binding result.

Furthermore, we investigate the confinement of electrons inside the strained region. It was shown earlier, using the Dirac equation formalism, that such a Gaussian bump results in low energy localized states\cite{M16,M17}. However, those models do not fully explain the origin of the six-fold symmetry of the localized states. Here we investigate the system using the tight-binding model and show that the influence of strain in the zigzag (zz) and armchair (ac) directions of graphene result in different pseudo-magnetic fields and consequently to different localization properties for the electrons. Furthermore, we examine the energy levels and wave functions in order to show the different confinement regimes.

This paper is organized as follows. In Sec. II we present the tight-binding model, the system geometry and the specific strain model that we use in the present paper. In Sec. III we evaluate the different approximations for the pseudo-magnetic field for high strain. In Sec. IV we calculate the electronic states using the tight-binding approach and we compare the results with the Landau levels predicted by the pseudo-magnetic field model. We also examine the confined electronic states inside the strained region. Our concluding remarks are given in Sec. V.

\section{The model}

\begin{figure}[h]
  \centering
  \includegraphics[width=8.6cm]{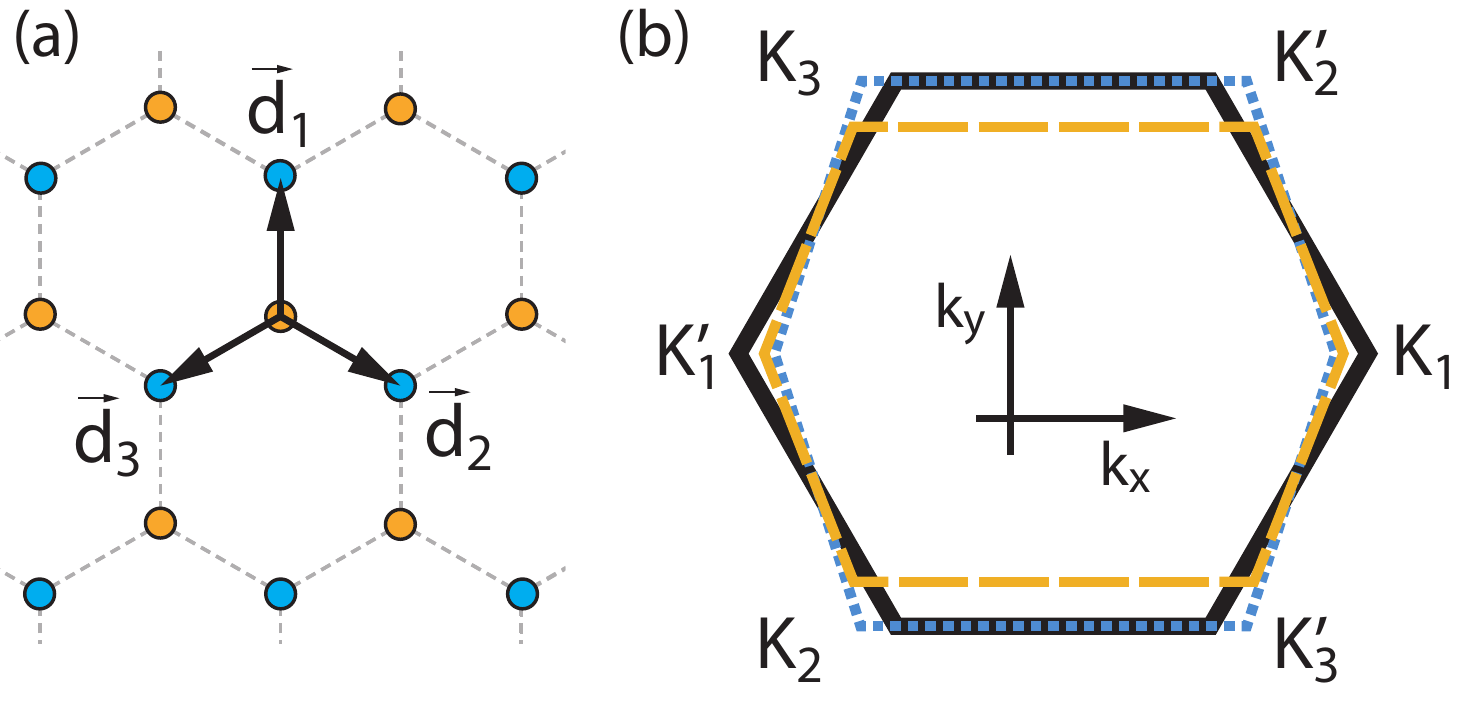}
  \caption{(Color online) (a) The unstrained nearest-neighbor vectors $\vec{d}_{n0}$. (b) The six $K$ points in the unstrained Brillouin zone (black, solid). The zone is also shown for $20\%$ armchair uniaxial strain, as calculated from the first approximation of the pseudo-magnetic vector potential (blue, dotted) and from the full solution of the vector potential (orange, dashed).  }
  \label{fig:Intro_2}
\end{figure}

We consider the tight-binding model of graphene with the nearest-neighbor Hamiltonian,
\begin{equation}\label{Hamiltionian}
     H = \sum_{m,n} t_{mn} a^{\dagger}_m b_n + h.c.
\end{equation}
Here $t_{mn}$ is the strained hopping energy between nearest-neighbor atoms at lattice positions $m$ and $n$, while $a_m$ and $b_n$ are field operators acting respectively on sublattices A and B at their given positions. Previously, it has been shown that the strained hopping parameter is given by\cite{M14},
\begin{equation} \label{strain_t}
    t_{mn} = t_0 \e^{-\beta \omega_{mn}},
\end{equation}
where $\omega_{mn} = l_{mn}/a_{cc} - 1$. Here $t_0 = -2.8$ eV is the unstrained hopping parameter, $l_{mn}$ is the strained distance between atoms $m$ and $n$, $a_{cc} = 0.142$ nm is the unstrained carbon-carbon distance and $\beta = 3.37$ is the strained hopping energy modulation factor. The nearest-neighbor vectors are $\vec{d}_1 = a_{cc} (0,1)$, $\vec{d}_2 = a_{cc}/2 (\sqrt{3},-1)$ and $\vec{d}_3 = a_{cc}/2 (-\sqrt{3},-1)$ as shown in Fig. \ref{fig:Intro_2}(a). The corresponding Brillouin zone and the six $K$-points are shown in Fig. \ref{fig:Intro_2}(b).

In the present paper we consider a finite size system which is taken as a hexagon with armchair edges. There are $N_{S}$ atoms on the hexagon edge, which corresponds to an edge width of $W_{S} = a_{cc} (3N_{S}/2  - 1)$. The total number of atoms in this hexagonal system is $N = 9 N_{S} \left( N_{S}/2 - 1 \right) + 6$. We limit ourselves to a hexagonal system that consists only of armchair edges in order to avoid the presence of zigzag edge states which would draw attention away from the bump-induced states. In the following calculations we take an edge width of $W_S=9$ nm, which corresponds with a flake consisting of 8322 C-atoms. The $x$-axis of the system is aligned with the zigzag direction in graphene.

In our model we strain the graphene flake by using a Gaussian bump located at the center of this system as illustrated in Fig. \ref{fig:Intro_1}(a). Such a strain profile can be induced with an STM tip\cite{Klimov2012}. The bump's height profile is given by $h(r) = h_0 \e^{-{r^2}/{b^2}}$, where $r$ is the distance from the center of the system, and $h_0$ and $b$ are parameters that characterize the Gaussian bump. The Gaussian function is defined to infinity ($r\rightarrow \infty$), which is inconvenient because increasing the system size would also change the total area of the bump. For that reason we add a cut-off radius $R$ after which the height of the bump will be zero. With this cut-off the bump height profile is expressed as,
\begin{equation}\label{gbump_limited}
    h(r) = h_0 \e^{-{r^2}/{b^2}} \Theta(R-r),
\end{equation}
where $\Theta$ is the Heaviside step function. It is important to choose the cut-off radius $R$ correctly in relation to the width parameter $b$ so that the most significant part of the bump is included before the cut-off. Taking $R = 3b/\sqrt{2}$ will ensure that $99.7\%$ of the Gaussian is inside the radius $R$. In the following calculations we take $R=6.2$ nm as typically realized in experiments\cite{M12}.

\section{The Pseudo-Magnetic Field}

The pseudo-magnetic vector potential in graphene $\vec{A}_{ps} = (\Re A_{ps}, \Im A_{ps})$ is given by\cite{M1},
\begin{equation} \label{A_ps_TBM}
    A_{ps} = \frac{1}{e v_F} \sum_{n=1}^3 t_n \e^{-i \vec{K} \cdot \vec{d}_n },
\end{equation}
where $\vec{d}_n$ and $t_n$ are the strained nearest-neighbor vectors and hopping parameters, respectively, and $\vec{K}$ is the location of a $K$ point. The pseudo-magnetic field is found as $\vec{B}_{ps} = \vec{\nabla} \times \vec{A}_{ps}$.

\begin{figure}[b]
  \centering
  \includegraphics[width=8.6cm]{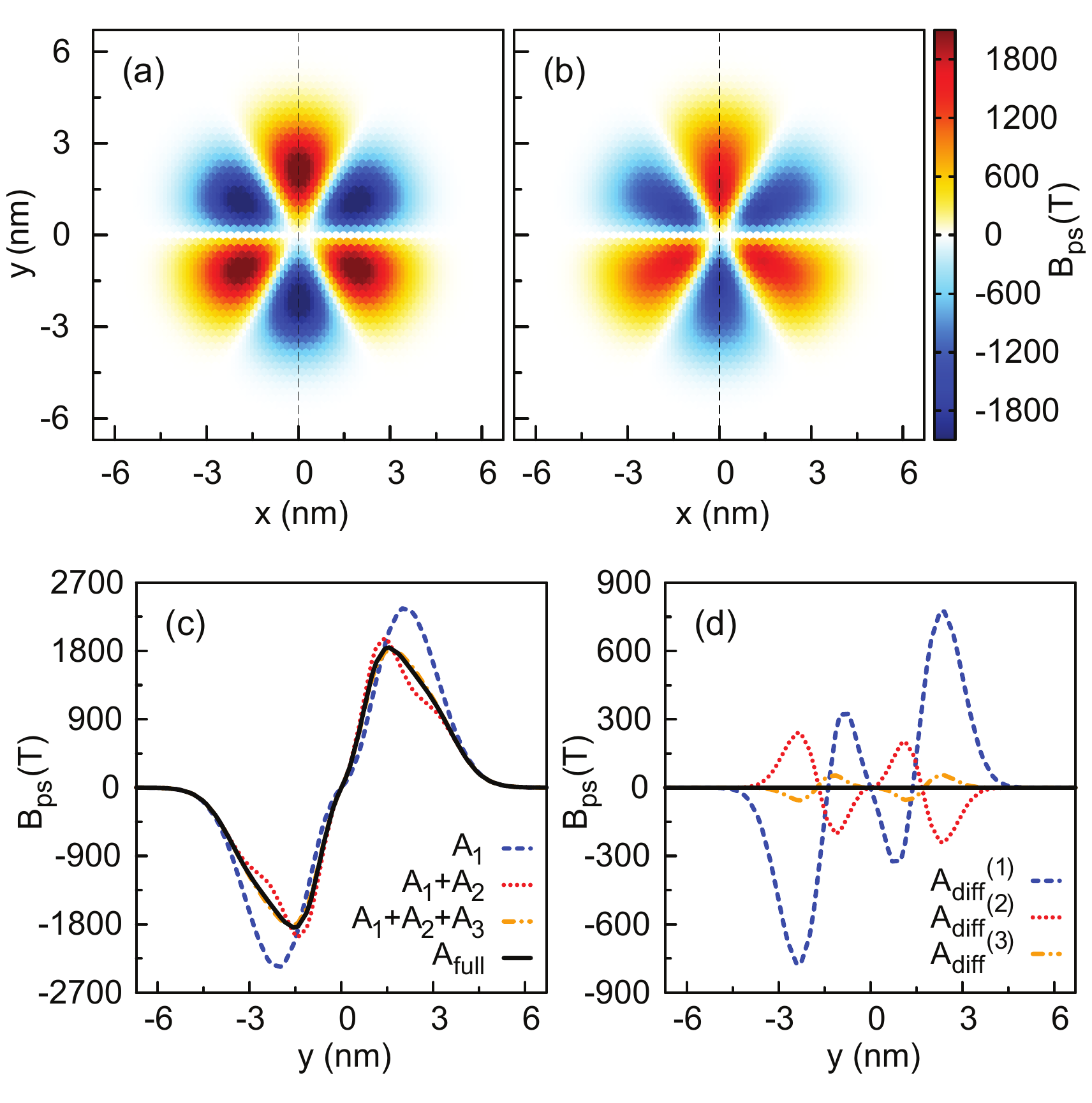}
  \caption{(Color online) Top: Contour plots of the pseudo-magnetic field generated by a Gaussian bump. The field is calculated (a) using the $A_1$ approximation of Eq. (\ref{A_ps_exp}) and (b) using the full form of the vector potential Eq. (\ref{A_ps_TBM}). The dashed lines show cuts at $x=0$, along the armchair direction of graphene. Bottom: (c) Plot of the field calculated using successively higher order terms of the vector potential approximation ($A_1$, $A_2$ and $A_3$) as well as the full form $A_{full}$ from Eq. (\ref{A_ps_TBM}). (d) The difference between the approximations and full solution as $A_{diff}^{(i)}=A_i-A_{full}$. In all cases the height of the bump is $h_0 = 2.2$ nm, which corresponds to a peak strain of $20\%$. }
  \label{fig:Bps_1}
\end{figure}

\begin{figure}[t]
  \centering
  \includegraphics[width=8.6cm]{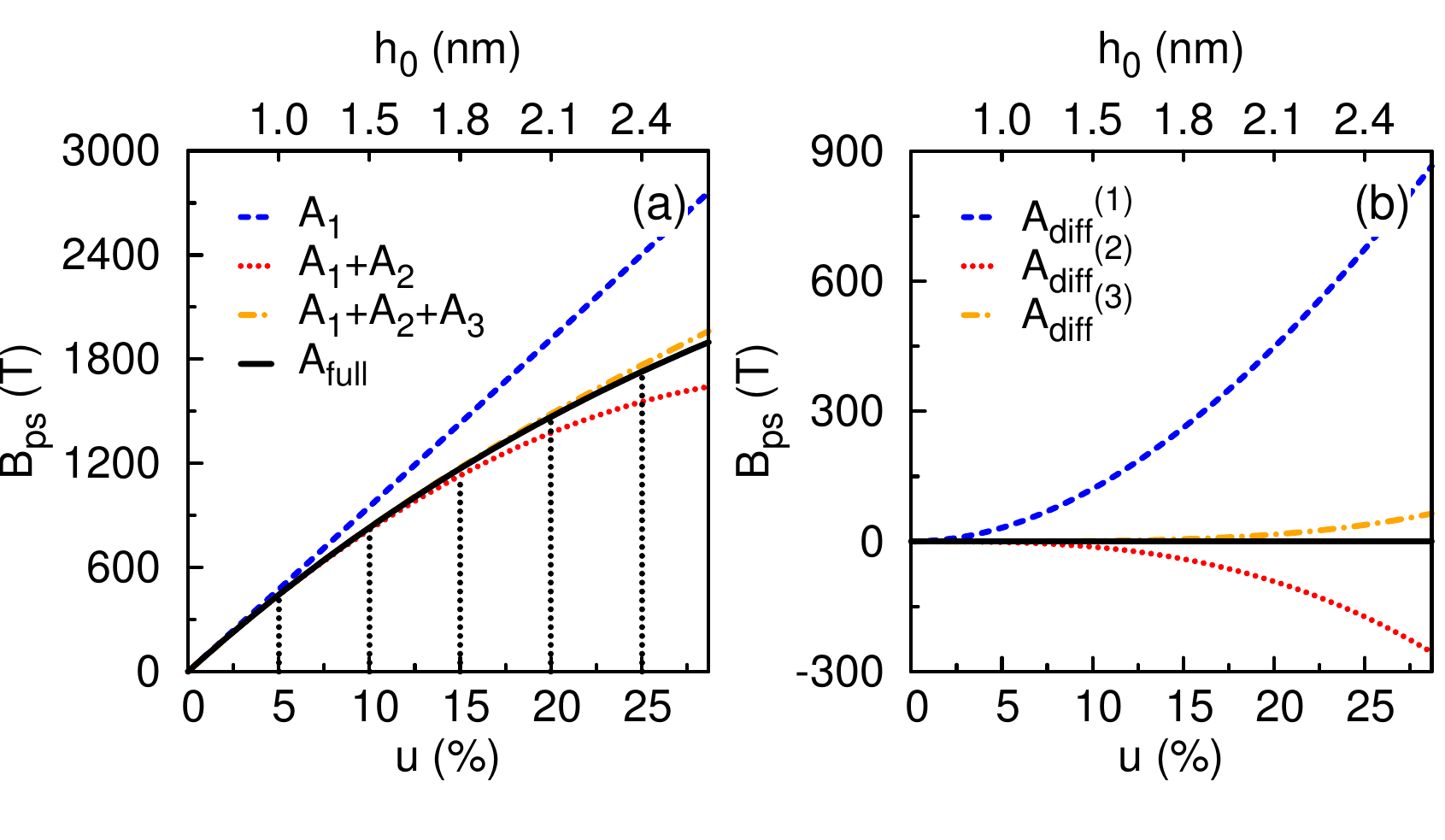}
  \caption{(Color online) (a) Pseudo-magnetic field at the location of maximum strain along the cut ($x=0$, $y=2.2$ nm). (b) The difference between the approximations and full solution as $A_{diff}^{(i)}=A_i-A_{full}$. The bump height $h_0$ is increased from 0 to $2.5$ nm, as indicated on the top $x$-axis, which generates the strain shown on the bottom $x$-axis.}
  \label{fig:Bps_2}
\end{figure}

The strained hopping parameter from Eq. (\ref{strain_t}) can be expanded to third order as,
\begin{eqnarray} \label{strain_t_exp}
    t_n &\approx&  t_0 + \delta t^{(1)}_n + \delta t^{(2)}_n + \delta t^{(3)}_n, \\
    t_n &\approx& t_0 \left(1 - \beta\omega_{n} + \frac{1}{2} \beta^2 \omega_{n}^2 - \frac{1}{6} \beta^3 \omega_{n}^3 \right).
\end{eqnarray}
The nearest-neighbor vectors $\vec{d}_n$ are also strained, but their total contribution to the pseudo-magnetic field is zero for any strain, so they may be safely neglected\cite{Kitt2012e,Juan2013}. While their inclusion would change the value of the vector potential, the resulting field would not be affected. As we are mainly interested in the pseudo-magnetic field, we will use the unstrained values of the vectors which are constant. Because of the out-of-plane deformation, the hopping will also be affected by curvature (hybridization between $\pi$ and $\sigma$ bands), but this contribution may be omitted as it is $100$ to $1000$ times smaller than the changes induced by the bond length modulation\cite{M1}.

Plugging the expansion (\ref{strain_t_exp}) into Eq. (\ref{A_ps_TBM}), we can expand the pseudo-magnetic vector potential to third order as,
\begin{equation} \label{A_ps_exp}
  A_{ps} \approx \frac{1}{e v_F} \sum_{n=1}^3 \bigg(
  \underbrace{ \delta t^{(1)}_n }_{A_1} +
  \underbrace{ \delta t^{(2)}_n }_{A_2} +
 \underbrace{ \delta t^{(3)}_n }_{A_3}
  \bigg) \e^{-i \vec{K} \vec{d}_{n}},
\end{equation}
which we subdivided into three parts $A_i$. $A_1$ is a first order term that was originally derived in Ref. \onlinecite{Suzuura2002}. $A_2$ and $A_3$ are second and third order terms which turn out to be important for large strain.

Fig. \ref{fig:Bps_1}(a) shows the pseudo-magnetic field calculated from the first approximation ($A_1$) of the vector potential. It exhibits three-fold symmetry with positive and negative peaks along the armchair directions of graphene and zero field along the zigzag directions. The pseudo-magnetic field based on the full vector potential, Eq. (\ref{A_ps_TBM}), without any approximations, is shown in Fig. \ref{fig:Bps_1}(b). To better see the difference in field magnitude between the different approximations, we take a cut along the armchair direction of graphene, as show in Fig. \ref{fig:Bps_1}(c). We compare the pseudo-magnetic field resulting from the vector potential approximations with successively higher terms included ($A_1$, $A_2$ and $A_3$) with the full form $A_{full}$ from Eq. (\ref{A_ps_TBM}). The differences are shown in Fig. \ref{fig:Bps_1}(d) as $A_{diff}^{(i)}=A_i-A_{full}$. The first order approximation $A_1$ overestimates the magnitude by as much as $800$ T. Adding the second order corrections ($A_2$) will give better agreement, but there are still large deviations in the region around $y=2.2$ nm where the strain is maximum, as well as near the center of the bump. Finally, including the third order term $A_3$ will result in generally good agreement.

In order to better evaluate the accuracy of the different vector potential approximations as a function of the strain, we plot the field at a fixed point while changing the bump height. As can be seen in Fig. \ref{fig:Bps_2}(a) and (b), approximation $A_1$ diverges from the full solution at values as low as $5\%$ strain. Adding $A_2$, we find good agreement up to about $15\%$, after which the field is increasingly underestimated. Finally, adding term $A_3$ yields good agreement up to $25\%$ strain.

These results bring up two issues. First, even the second order term $A_2$ is not enough to sufficiently approximate the pseudo-magnetic field for strain above $15\%$. Expanding the approximation to third order would improve results, but that would just needlessly complicate matters. Second, even if the second order term were sufficiently accurate, it's form is too complicated for analytical results. On the other hand, using the numerical approach, there is no need for this, as the full vector potential Eq. (\ref{A_ps_TBM}) can easily be calculated. Thus, we find that numerical methods are best suited for correctly calculate the pseudo-magnetic field at graphene's high tolerance of up to $25\%$ strain.

\section{The electronic states}

We derive the energy levels $E_n$ and wave functions $\Psi(x,y)$ of the bump strained graphene flake using the tight-binding Hamiltonian Eq. (\ref{Hamiltionian}) with the effect of strain included via the modulation of the hopping parameter given by Eq. (\ref{strain_t}). We shall compare the results from the tight-binding approach with the pseudo-magnetic field model from the previous section.

\begin{figure}[t]
  \centering
  \includegraphics[width=8.6cm]{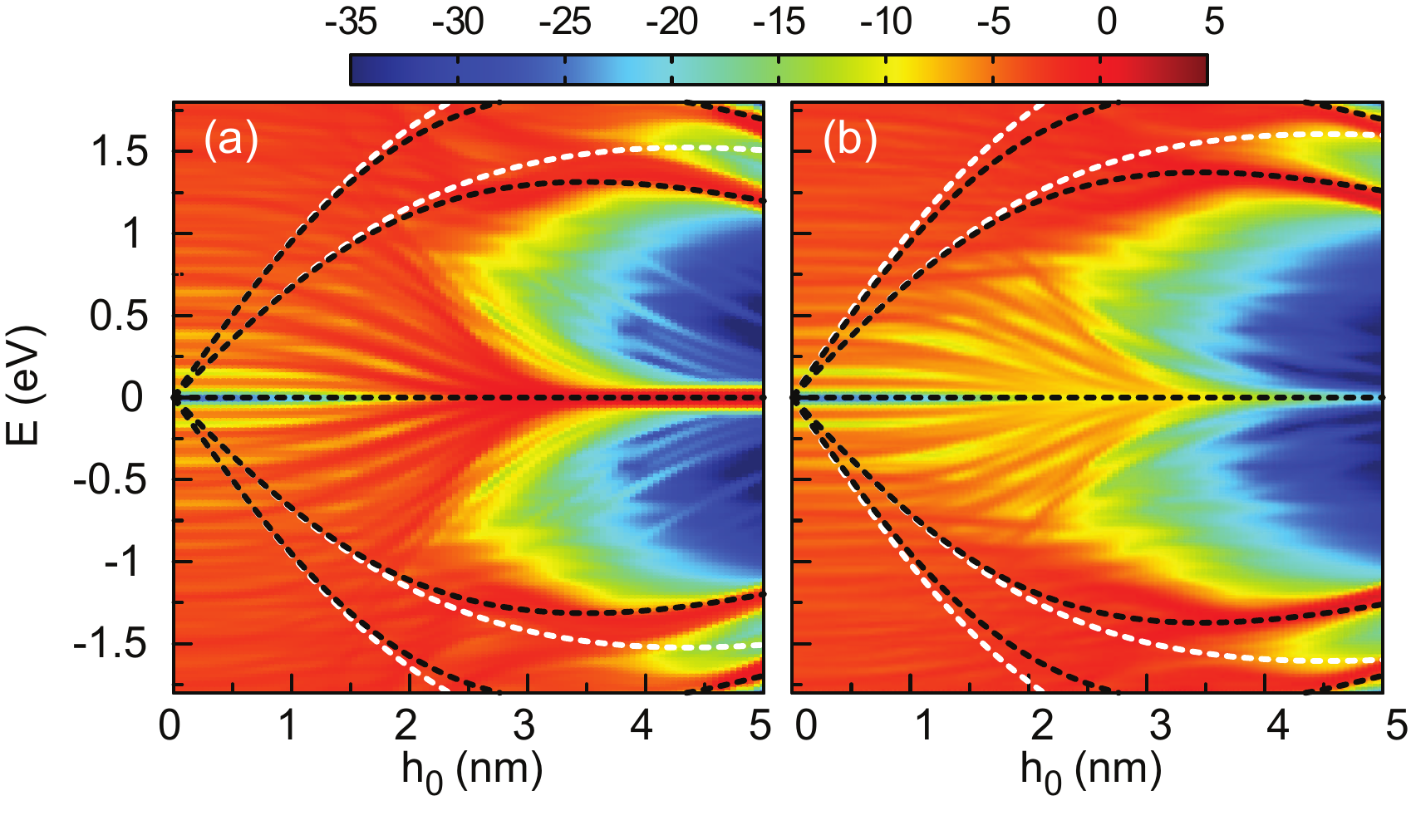}
  \caption{(Color online) Contour plot of the LDOS for sublattice (a) A and (b) B as a function of bump height and energy, at the location of maximum strain ($x=0$, $y=2.2$ nm). The dashed curves are the Landau levels based on the pseudo-magnetic field model, calculated using unstrained (white curves) and strained (black curves) Fermi velocity. }
  \label{fig:Bps_3}
\end{figure}

The local density of states (LDOS) is given by,
\begin{equation}\label{LDOS}
    D(E,x,y) = \sum_{n} |\Psi(x,y)|^2 \delta(E - E_n).
\end{equation}
To calculate the LDOS numerically we introduce a Gaussian broadening,
\begin{equation}
    \delta(E-E_n) \rightarrow \frac{1}{\Gamma \sqrt{\pi}} \exp{\left[-\frac{(E -
    E_{n})^2}{\Gamma^{2}}\right]}.
\end{equation}
As we did previously in Fig. \ref{fig:Bps_2}, we select the location of highest strain along the armchair direction and we calculate the LDOS at that point in space as a function of bump height and energy. The results are shown in Fig. \ref{fig:Bps_3}. At large bump heights, the LDOS shows Landau levels up to the second. Sublattice B has a lower LDOS at the zero Landau level, which is consistent with what is found for a magnetic field in graphene.

For comparison, we calculate the Landau levels using the pseudo-magnetic field model from the previous section and we overlay them on the LDOS as dashed curves in Fig. \ref{fig:Bps_3}. In this case the pseudo-magnetic field is calculated according to the full vector potential Eq. (\ref{A_ps_TBM}). Note that the Landau levels do not follow the usual square root function. This is because the plot is a function of bump height. The Landau levels still behave as a square root of the pseudo-magnetic field.

In the first case (white dashed curves), the Landau levels are plotted for a constant unstrained Fermi velocity $v_F = 3a_{cc} t_0 / 2 \hbar$. This does not give good agreement with the LDOS when the bump height is large. In the second case (black dashed curves), the Landau levels are fitted to the LDOS with an adjusted strained Fermi velocity $v_F^{(s)} = 3 (a_{cc} t_0  + \alpha l_{mn} \delta t_{mn}) / 2 \hbar$, where $\delta t_{mn} = t_{mn}-t_0$, and $\alpha = 0.28$ is a fitting constant.

\begin{figure}[b]
  \centering
  \includegraphics[width=8.6cm]{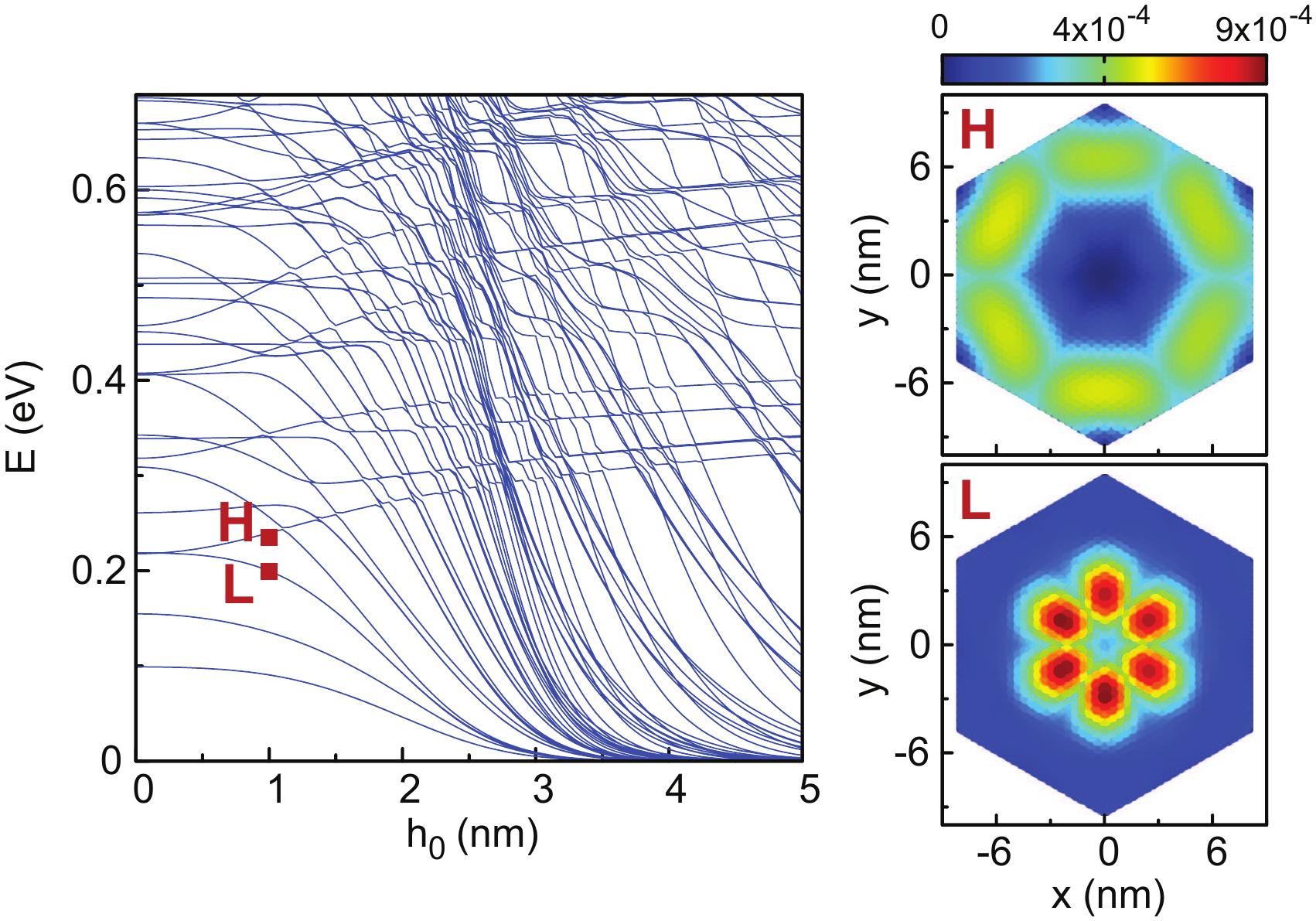}
  \caption{(Color online) Energy spectrum of a hexagonal armchair graphene flake strained by a Gaussian bump. Left: Energy levels as a function of bump height. Right: Spatial probability at the split levels marked with $L$ and $H$.}
  \label{fig:States_1}
\end{figure}

Next, we are interested in finding the spatial localization of the different electron states. We plot the energy levels as a function of bump height in Fig. \ref{fig:States_1}. The levels split into two groups: some energy levels decrease toward zero as the height of the bump increases, while the other group has the opposite behavior and increases slowly in energy with $h_0$. To better understand these two types of levels, we examine their wave functions. A point on a rising energy level is marked as $H$ in Fig. \ref{fig:States_1}. The spatial probability for this state (see the right panel in Fig. \ref{fig:States_1}) shows an electron state localized away from the center of the system, i.e. it is localized outside the bump. As the height of the bump is increased, the confinement area between the bump and the system edge is reduced. This reduction in confinement area results in an increase of the energy.

\begin{figure}[t]
  \centering
  \includegraphics[width=8.6cm]{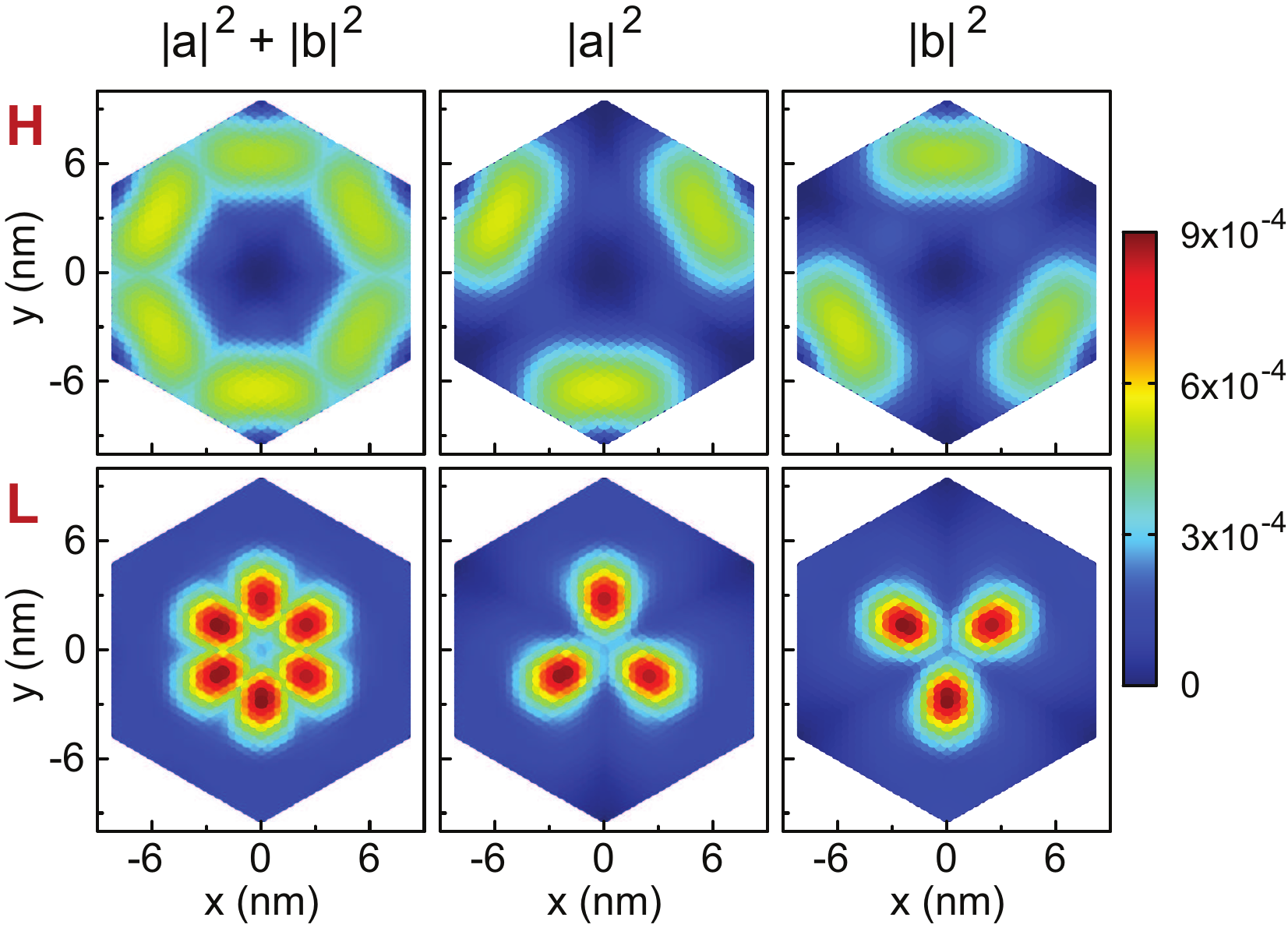}
  \caption{(Color online) Electron probability for different sublattices at the points $L$ (bottom figures) and $H$ (top figures) from Fig. \ref{fig:States_1}.}
  \label{fig:States_2}
\end{figure}

\begin{figure}[b]
  \centering
  \includegraphics[width=8.6cm]{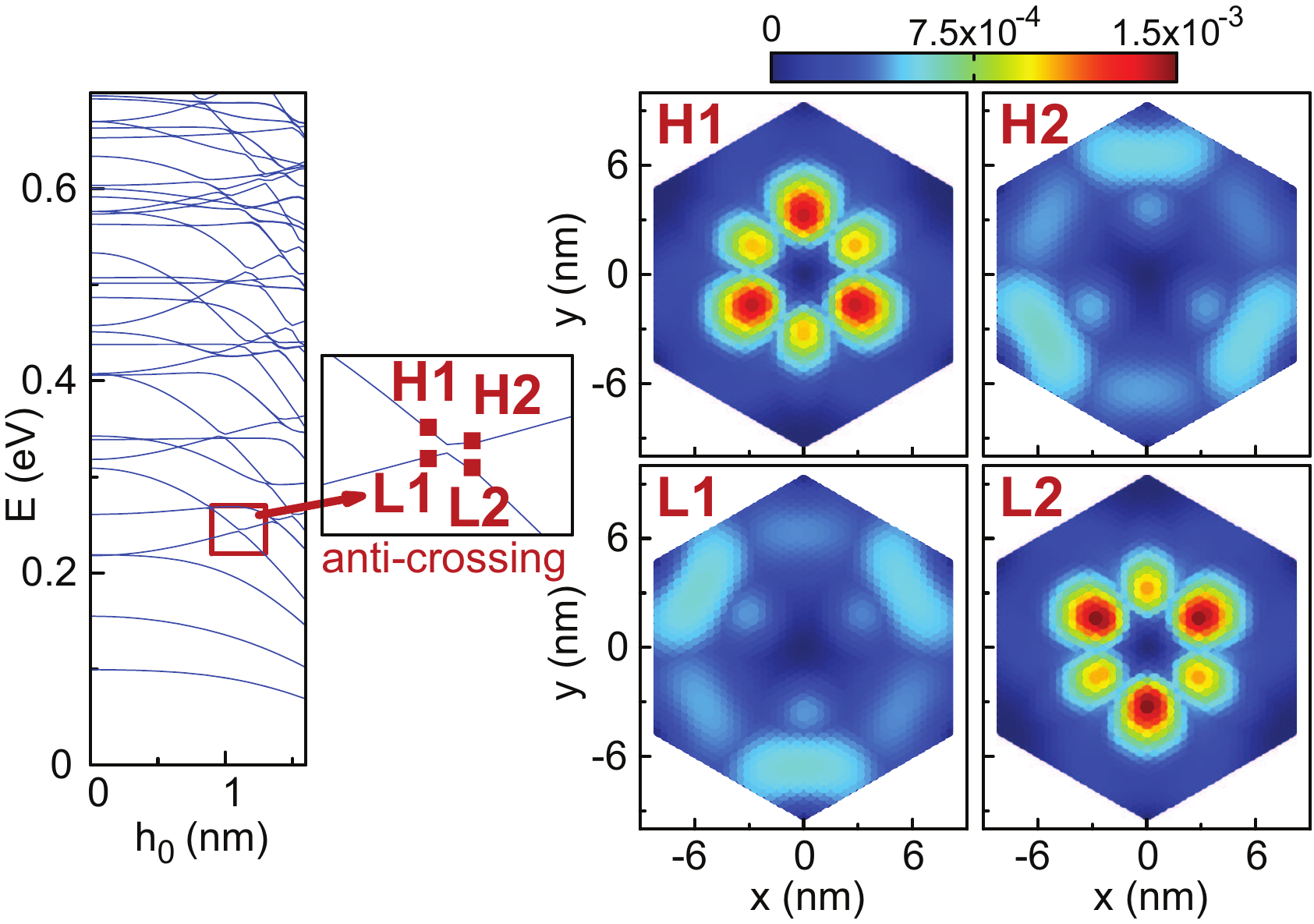}
  \caption{(Color online) Energy levels as a functions of bump height in a hexagonal graphene flake with armchair termination. Left: Anti-crossing point at bump height $h_0 = 1$ nm and energy $E=0.24$ eV. Right: Spatial probability before and after the anti-crossing point.}
  \label{fig:States_3}
\end{figure}

On the other hand, we have point $L$ which marks the level that splits downward, away from level $H$. The spatial probability in point $L$ is plotted in the right bottom panel of Fig. \ref{fig:States_1}. These decreasing levels are confined inside the bump, in contrast to the previous case. The probability peaks are found in the armchair directions which coincides with the peaks of the pseudo-magnetic field from Fig. \ref{fig:Bps_1}. These levels converge toward zero energy, thus forming the zero Landau level. Because the pseudo-magnetic field is nonhomogeneous in this system, higher Landau levels are not clearly visible in the global energy spectrum.

\begin{figure}[t]
  \centering
  \includegraphics[width=8.6cm]{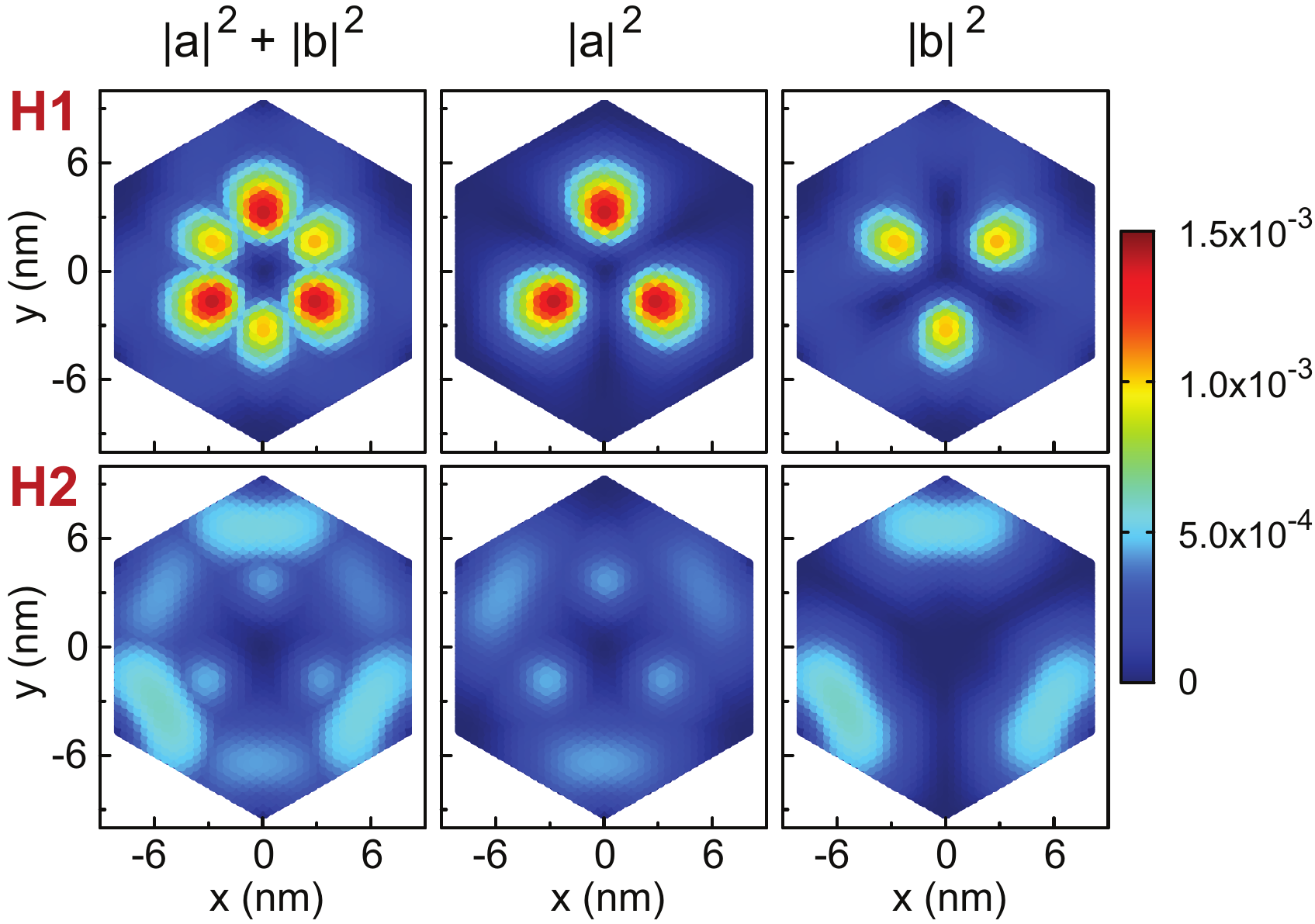}
  \caption{(Color online) Electron probability for different sublattices before and after the anti-crossing point from Fig. \ref{fig:States_3}.}
  \label{fig:States_4}
\end{figure}

\begin{figure}[b]
  \centering
  \includegraphics[width=8.6cm]{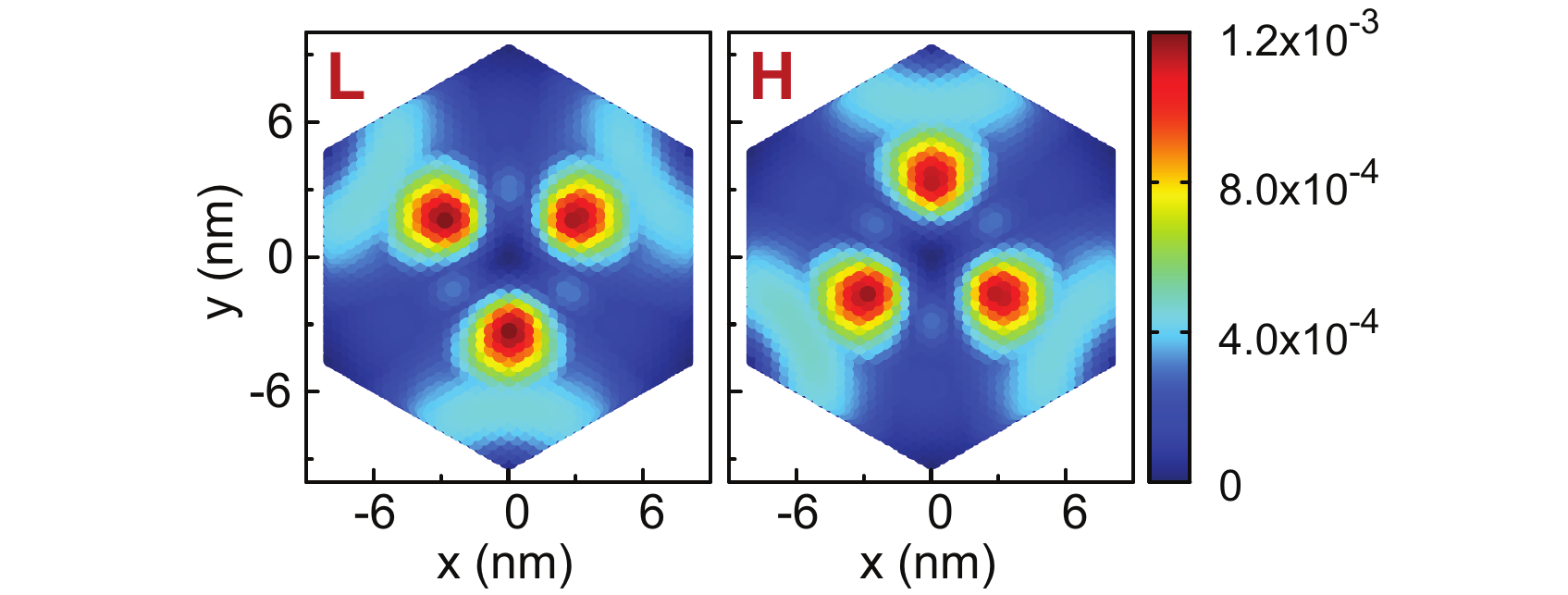}
  \caption{(Color online) Electron probability at the lower (L) and higher (H) extrema of the anti-crossing shown in Fig. \ref{fig:States_3}.}
  \label{fig:States_5}
\end{figure}

The wave functions on the individual sublattices A and B are shown in Fig. \ref{fig:States_2} for both the $L$ and $H$ branches. Each sublattice contributes to three of the six probability peaks. The peak heights for each sublattice are the same, but the peaks are positioned opposite to one another. The areas of high probability for sublattice A are positioned at the positive peaks of the pseudo-magnetic field, while those localized on sublattice B coincides with the negative peaks of the pseudo-magnetic field. Note that for the $H$ level, the probability peaks are rotated by $60^\circ$ in the two sublattices as compared to the $L$ level. This also points to the different origin of these levels.

Another interesting feature of the energy levels are anti-crossing points that switch the two types of energy levels. We examine one of these anti-crossing points in detail in Fig. \ref{fig:States_3}. We mark points $H1$, $H2$ ($L1$, $L2$) on the higher (lower) level before and after the anti-crossing, respectively. Following the higher level from $H1$ to $H2$, we can see a transition from confinement inside the bump to confinement outside the bump. This is consistent with the previously discussed confinement types for decreasing and increasing energy levels with $h_0$. Following the lower level from $L1$ to $L2$ reveals the opposite behavior, with the confinement switching from outside to inside. Note also that when we go from $H1$ ($L1$) to $L2$ ($H2$) the position of the peaks are rotated by $60^\circ$. The direction of the appearance of the peaks in the lower (higher) branch does not change when passing the anti-crossing point.

\begin{figure}[t]
  \centering
  \includegraphics[width=8.6cm]{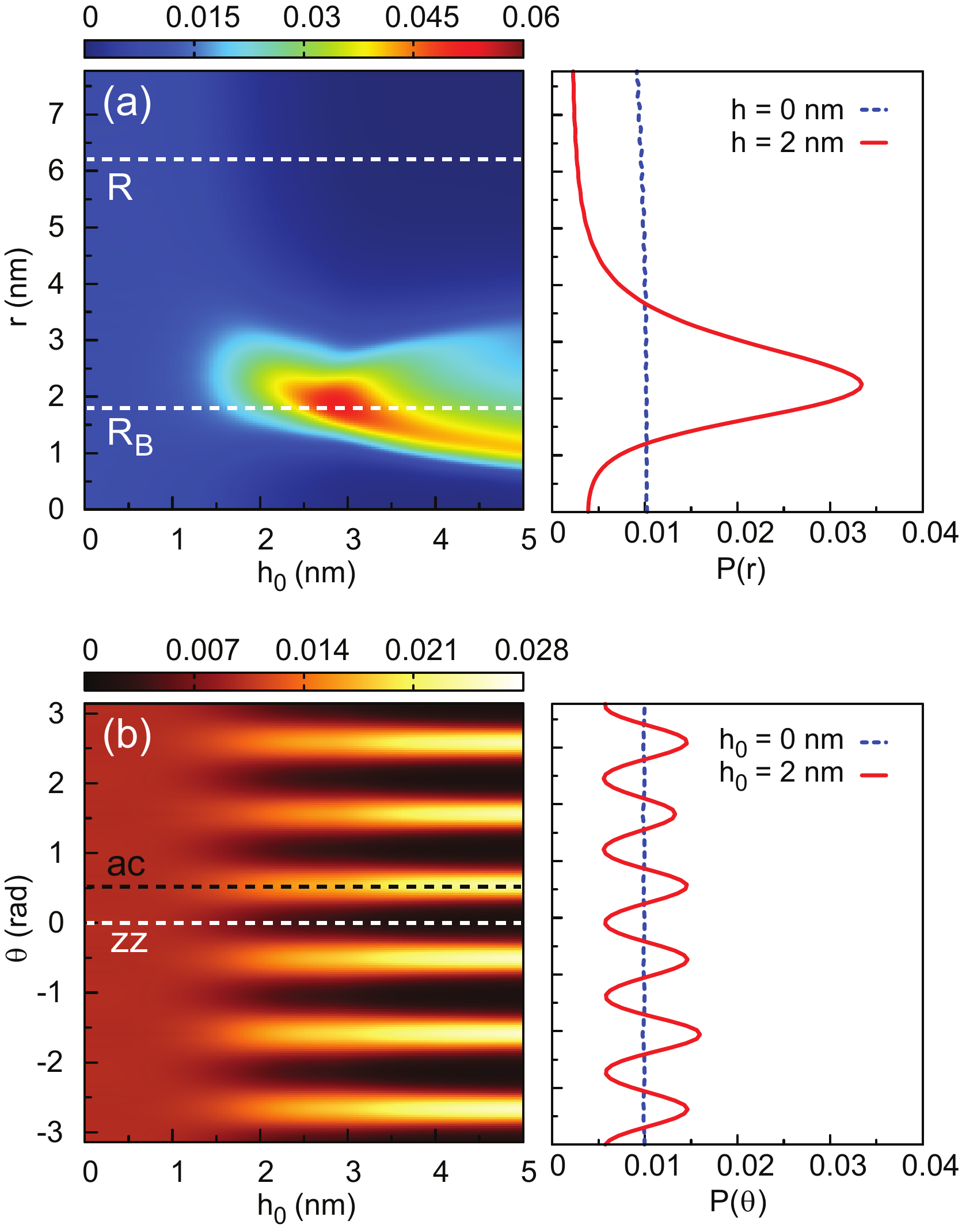}
  \caption{(Color online) Left: (a) Radial and (b) angular electron probability at low energy as a function of the bump height. The dashed line labeled $R$ marks the radius of the bump and $R_B$ is the spatial position of the maximum of the pseudo-magnetic field. In figure (b) the dashed lines indicate the armchair and zigzag directions in graphene. Right: Cuts of the probability at $h_0 = 0$ and $h_0 = 2$ nm. The energy is fixed at $E = 0.05$ eV.}
  \label{fig:Conf_1}
\end{figure}

\begin{figure}[t]
  \centering
  \includegraphics[width=8.6cm]{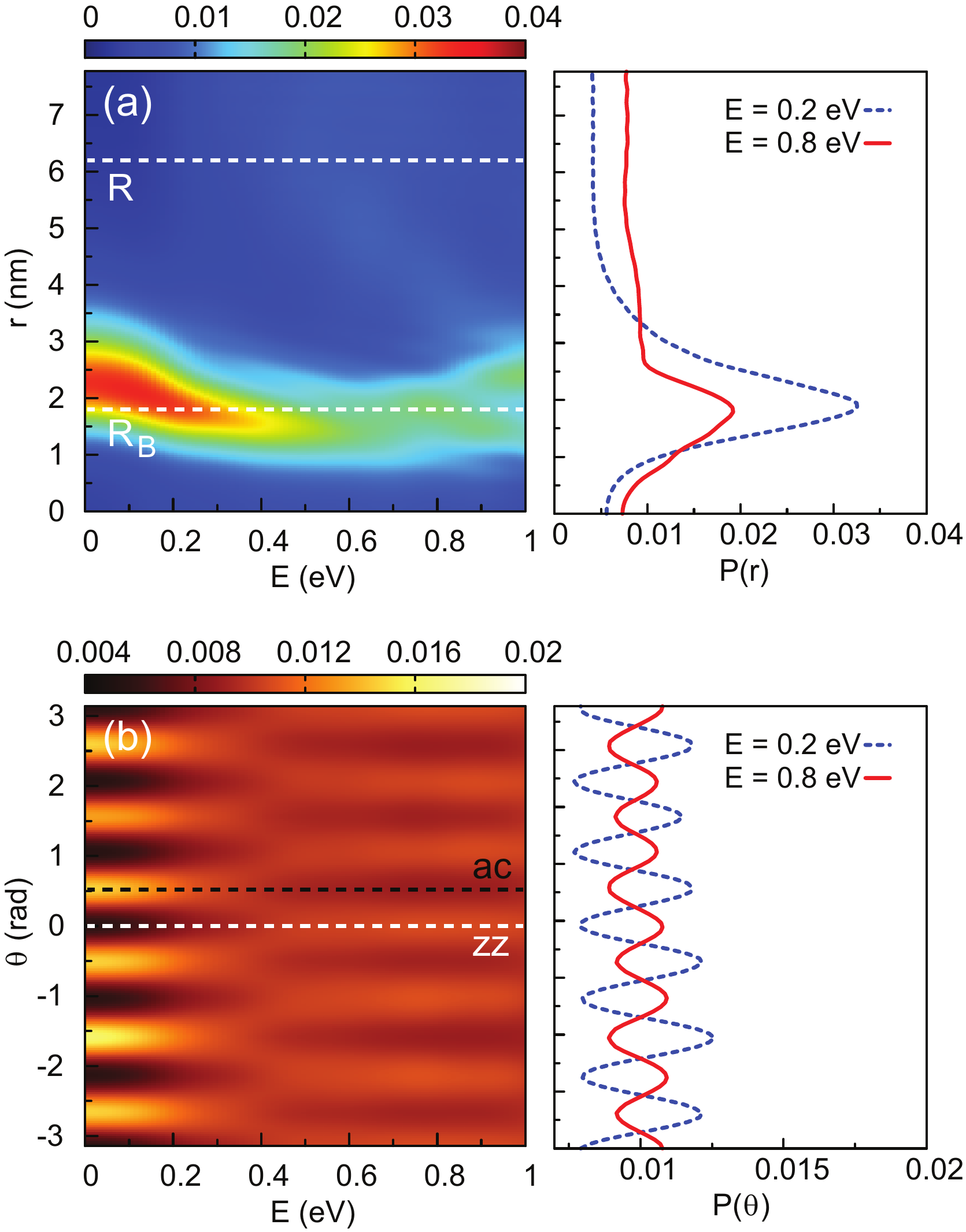}
  \caption{(Color online) Left: (a) Radial and (b) angular electron probability as a function of energy. The marked lines are the same as in Fig. \ref{fig:Conf_1}. Right: Cuts of the probability at $E=0.2$ eV and $E=0.8$ eV. The bump height is fixed at $h_0 = 2$ nm.  }
  \label{fig:Conf_2}
\end{figure}

The probability plots around the anti-crossing point do not show perfect six-fold symmetry as we have seen in previous cases. Instead, we have two sets of three probability peaks with different magnitudes. We examine this asymmetry in Fig. \ref{fig:States_4} by plotting the separate probabilities for the two sublattices. At point $H1$, sublattice A has larger probability, but at point $H2$ (after the anti-crossing) this is reversed. Thus, following an energy level through an anti-crossing point from $H1$ to $H2$ (or $L1$ to $L2$) will result in two transitions: both the confinement type (inside or outside the bump) and the sublattice dominance are switched.

For completeness, we plot the probability distribution at the extrema of the anti-crossing in Fig. \ref{fig:States_5}. Notice that they exhibit an appreciable probability both inside and outside the bump. Both points are three-fold symmetric but rotated by $60^\circ$ relative to one another.

Since the bump is radial, it is natural to express the electron probability in polar coordinates as $P(r, \theta)$. We are specifically interested in finding electron states that are confined inside the bump and that are not influenced by the finite size of the simulated system. We can take an integral over the angle and only leave the radius dependent part of the probability
\begin{equation}
    P(r) = \int_0^{2\pi} P(r,\theta) d \theta.
\end{equation}
Alternatively, we can do the opposite and take the integral over the radius, which leaves just the dependence on the angle.

The radial probability of low energy electrons as a function of bump height is shown in the left part of Fig. \ref{fig:Conf_1}(a). As expected, the probability near zero bump height is practically uniform across the full radius of the system. As the height of the bump increases, we start seeing confinement inside the bump. More specifically, the probability peak is close to the position of the maximum of the pseudo-magnetic field, marked as $R_B$.

Fig. \ref{fig:Conf_1}(b) shows the angular probability. The dashed lines indicate the armchair (black) and zigzag (white) directions of graphene, which alternate every $\pi/6$ radians. For low bump height, the angular distribution is practically uniform. As the height of the Gaussian bump increases, probability maxima start to form in the armchair directions and minima appear in the zigzag directions.

Next, we fix the bump height at $h_0=2$ nm and we present the probability as a function of energy in Fig. \ref{fig:Conf_2}(a). For low energy, we find that the state is mostly confined inside the bump near the strain maximum. But for energies above $0.3$ eV, we find substantial probability outside the bump and thus weaker confinement.

\begin{figure}[t]
  \centering
  \includegraphics[width=8.6cm]{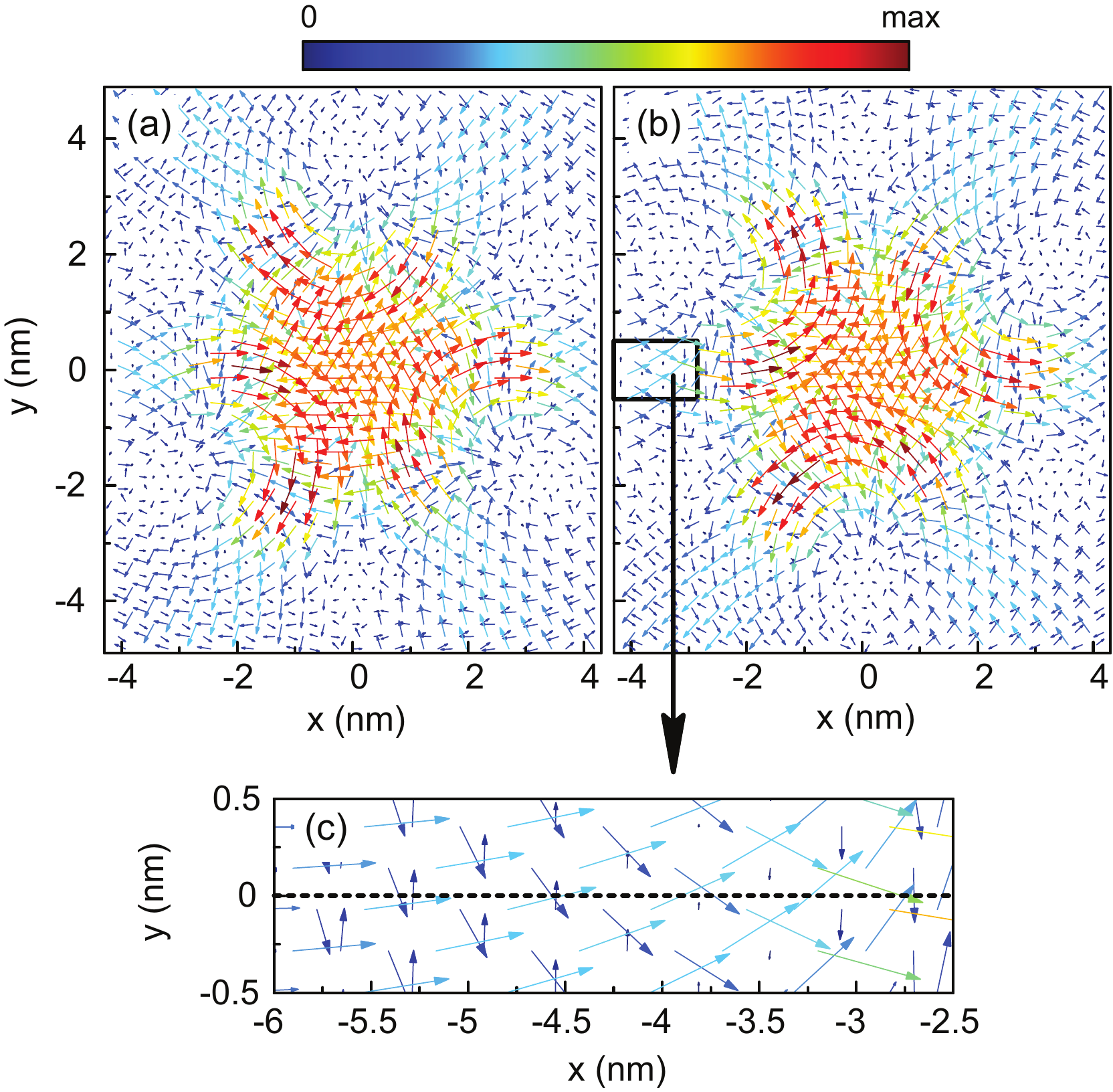}
  \caption{(Color online) Probability current for the $L$ point electron state from Fig. \ref{fig:States_1}, for sublattices (a) A and (b) B. (c) An enlarged region around snake states. The dashed line indicates the zigzag direction. }
  \label{fig:Current_1}
\end{figure}

Looking at the angular plot in Fig. \ref{fig:Conf_2}(b), we find probability peaks in the armchair directions and minima in the zigzag directions. As the energy increases the peaks disappear around $0.3$ eV, which is the same energy where we started seeing substantial probability outside the bump in the radial plot. Once the probability outside the bump becomes substantial (above $0.4$ eV), the highest probability shifts to the zigzag directions. As we will see later, the zigzag directions are associated with directions along which the probability current flows, connecting the center and the outside of the bump.

The single valley probability current at carbon atom $m$ is given by,
\begin{equation}\label{P_current}
  \vec{j}_m = \frac{i}{\hbar} \sum_{n=1}^3 \Psi_m^* H_{m,m+n} \Psi_{m+n} \vec{d}_n,
\end{equation}
where $H_{m,m+n} = -t_{m,m+n}$ is the tight-biding Hamiltonian matrix element. We plot the current inside the hexagonal flake in Fig. \ref{fig:Current_1} for the same $L$ state as in Fig. \ref{fig:States_1}. For clarity, the current is plotted separately for sublattices A and B. Circular orbits coincide with the positions of the six probability peaks in the armchair directions. The current is very low at the exact positions of the probability peaks, but there is an appreciable circular current flowing around the peaks. Each sublattice contributes three circular orbits, where the sublattice A orbits have a clockwise direction and sublattice B is counterclockwise. This coincides with the probability peaks of the individual sublattices, as well as with the positive and negative peaks of the pseudo-magnetic field. Lines where the pseudo-magnetic field is zero lie along the zigzag directions. Fig. \ref{fig:Current_1}(c) shows the current along this line, where we find the current flowing successively across the line in both directions. These are snake orbits which are present because of the interface between the positive and negative pseudo-magnetic field.

\section{Conclusions}

We showed that a circular symmetric straining of a hexagonal graphene flake induces a non-circular symmetric pseudo-magnetic field. The average induced pseudo-magnetic field is zero and the field changes sign when we cross a zigzag direction. The pseudo-magnetic field was calculated up to third order in the strain. The first order term was found to be valid only up to $5\%$ strain. The second order term extends the validity of the pseudo-magnetic field expression for a Gaussian bump up to $15\%$, while the third order is needed to go up to graphene's full strain limit ($25\%$).

Next, we investigated the confinement of electronic states in the same system. We found that non-uniform strain has a significantly different effect in the two fundamental directions of graphene. The six-fold symmetry of the confinement is directly related to the armchair and zigzag directions. Electrons are well confined in the armchair directions, while the zigzag directions allow the flow of probability current between the inside and outside of the strained region. This mirrors the form of the strain-generated pseudo-magnetic field, which has peaks in the armchair direction and zero magnitude in the zigzag directions.

The energy levels of the Gaussian bump system show splitting and anti-crossing states that correspond to different regions of localization of the electron. The levels that increase in energy with increasing bump height are confined between the bump and the edges of the graphene flake, while the decreasing levels correspond to states confined inside the bump. We also identified several anti-crossing points which switch the confinement type (inside or outside the bump) as well as the sublattice dominance.

We examined the probability of finding the electron in the system as a function of the radius and angle. At low energy there is strong confinement inside the bump near the strain maximum. At higher energy, confinement is weaker and as more states are found outside the bump. As for directional confinement, we found that low energy states are well confined in the armchair directions, where we see closed circular electron orbits. Higher energy states are more likely found in the zigzag directions, where the probability current shows channels to and from the center of the bump.

\section{Acknowledgment}

This work was supported by the European Science Foundation (ESF) under the EUROCORES Program EuroGRAPHENE within the project CONGRAN, the Flemish Science Foundation (FWO-Vl) and the Methusalem Funding of the Flemish Government.

\end{document}